\documentclass[aip,rsi,reprint]{revtex4-1} % for checking your page length

\usepackage{siunitx}
\DeclareSIUnit{\sqrthz}{\ensuremath{\sqrt{\text{\hertz}}}}

\usepackage{amsmath}
\usepackage{amsfonts}

\usepackage{graphicx}
\usepackage[caption=false]{subfig}
\usepackage{dcolumn}
\usepackage{multirow}
\usepackage{hyperref}

\DeclareGraphicsExtensions{.png,.tiff,.ai,.eps}
\graphicspath{{fig/}}

\newcommand{\prb}{Phys. Rev. B}
\newcommand{\iu}{\mathrm{i}\mkern1mu}

\usepackage{natbib}

\begin{document}

\title{An ultra-low noise, high-voltage piezo driver}

\author{N.C. Pisenti}
\email[]{npisenti@umd.edu}
\author{A. Restelli}
\author{B.J. Reschovsky}
\author{D.S. Barker}
\author{G.K. Campbell}
\affiliation{Joint Quantum Institute, University of Maryland and National Institute of Standards and Technology \\ College Park, MD 20742, USA}

\date{\today}

\begin{abstract}
We present an ultra-low noise, high-voltage driver suited for use with piezoelectric actuators and other low-current applications. 
The architecture uses a flyback switching regulator to generate up to 250V in our current design, with an output of \SI{1}{\kilo\volt} or more possible with small modifications. 
A high slew-rate op-amp suppresses the residual switching noise, yielding a total RMS noise of $\approx\SI{100}{\micro\volt}$ (\SI{1}{\hertz}--\SI{100}{\kilo\hertz}).
A low-voltage (\SI{\pm 10}{\volt}), high bandwidth signal can be summed with unity gain directly onto the output, making the driver well-suited for closed-loop feedback applications.
Digital control enables both repeatable setpoints and sophisticated control logic, and the circuit consumes less than \SI{150}{\milli\ampere} at $\pm\SI{15}{\volt}$.
\end{abstract}


\maketitle %\maketitle must follow title, authors, abstract and \pacs

\section{Introduction}
\label{Sec:Introduction}

Many instrumentation applications in the modern laboratory require agile, low-noise voltage sources capable of supplying hundreds of volts or more.
For example, piezo-actuated mirrors and diffraction gratings play an important role in atomic physics experiments (used, e.g., in Fabry-P\'{e}rot cavities\cite{Riedle1994a,Bohlouli-Zanjani2006a} and external-cavity diode lasers\cite{Wieman1991a}), while avalanche photodiodes and photomultiplier tubes require large bias voltages for proper operation.
In the realm of biophysics, electrokinetic separation methods such as free-flow or capillary electrophoresis\cite{Kohlheyer2008a} require large electric field gradients, and the recent push to develop lab-on-a-chip devices could benefit from miniaturized high-voltage sources.\cite{Temiz2015a}

Laboratory devices are often operated in a closed feedback loop, where small voltage changes on top of a large DC voltage are necessary to stabilize the output of a particular system.
For example, the frequency of an extended-cavity diode laser can be locked by feeding back to a piezo-actuated diffraction grating or mirror, which in turn supplies optical feedback to the diode.
Commercially available piezoelectric drivers typically provide a modulation input for such closed-loop applications, but the input voltage is often gained such that it spans the entire output range of the device.
Other designs separate high- and low-voltage control pathways, which can extend the bandwidth to $\approx\si{\mega\hertz}$, but the low-voltage control is AC-coupled to the output.\cite{Fleming2009a}
While these designs have advantages, many applications would benefit from an architecture that provides a unity-gain, DC-coupled feedback path to the high-voltage output.
This low-gain modulation input could make closed-loop systems less susceptible to noise contributions from the servo controller, which we often find in our laboratory to be a limiting factor in laser lock stability.

Instrumentation electronics capable of supplying high voltages traditionally fall under one of two architectural umbrellas: DC-DC switching converters, and linear-type amplifiers.
While DC-DC converters are efficient and can work at very high voltages, they suffer from switching noise and limited control bandwidths.
Linear-type devices are typically constructed from a high-voltage operational amplifier (op-amp), powered either from a high-voltage linear regulator or more typically from a secondary switching converter.
While the op-amp can provide \SI{100}{\decibel} or more of power-supply noise rejection\cite{PA84Datasheet}, linear regulators must handle any excess voltage by dissipating heat and so may be more cumbersome to deploy in the laboratory.

We present a circuit with a hybrid architecture.
The high voltage is generated by a galvanically isolated DC-DC converter, while a low-noise, high-slew-rate op-amp simultaneously removes noise at the output and provides a low-gain, high-bandwidth ($\gtrsim \SI{100}{\kilo\hertz}$) modulation input for closed-loop feedback applications.
This architecture is able to achieve extremely low noise ($\approx\SI{100}{\micro\volt}_{\text{RMS}}$) over the entire output range, draws very little current, and fits comfortably onto a small-footprint printed circuit board (PCB).
Additionally, the high-voltage output remains single-ended and referenced to ground, allowing it to drive piezo actuators with a grounded terminal.
The schematic is presented in Sec.~\ref{Sec:Circuit}, with a noise analysis in Sec.~\ref{Sec:NoiseAnalysis} and characteristic performance data in Sec.~\ref{Sec:Results}.
Complete design files, including the schematic, bill of materials, and board layout, can be found on GitHub.\cite{PiezoDesignFiles}
The board manufacture and component cost is less than \SI{200}[\$]{}, making it a cost-effective alternative to commercial options.

\section{Circuit Design}
\label{Sec:Circuit}

\begin{figure*}[t]
\includegraphics[width=\textwidth]{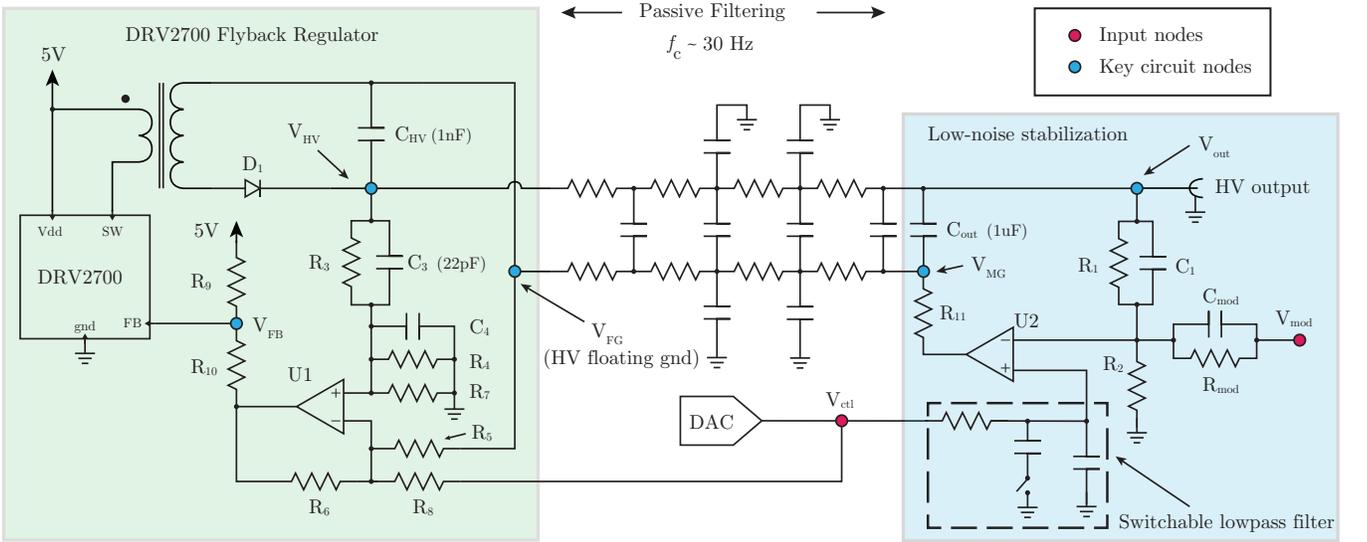}
\caption{Schematic of the high voltage supply and stabilization.
The voltage $\text{V}_\text{HV}$ is generated using a Texas Instruments DRV2700 high voltage driver in flyback configuration.
A high slew-rate op-amp (U2) senses the output voltage across $R_1$ and $R_2$, and controls it by modulating the mid-ground node $\text{V}_\text{MG}$.
The DC control signal for this op-amp,  $\text{V}_{\text{ctl}}$, is supplied by a digital-to-analog (DAC) converter, which is passed through a switchable low-pass filter. 
This design allows for very heavy filtering of the DAC $1/f$ noise during steady-state operation, but the corner frequency can be increased if the output needs to be scanned more quickly.
The $\text{V}_{\text{ctl}}$ gain is set by $\left(1+R_1/R_2\right)$, while the modulation gain is set by $-R_1/R_\text{mod}$.
The op-amp U2 removes residual switching noise and stabilizes the DC output according to the transfer function given in Eq.~(\ref{Eq:FullTransferFunc}).
A MOSFET quench circuit, shown in Fig.~\ref{Fig:MOS}, connects at nodes $\text{V}_\text{out}$ and $\text{V}_\text{MG}$.
\label{Fig:PiezoCircuit}}
\end{figure*}

The design principles discussed below show how we leverage the characteristics of a galvanically-isolated switching regulator without sacrificing the low-noise requirements of many laboratory applications.
Our design targets a \SI{250}{\volt} output, but we discuss straightforward modifications to the schematic that make it possible to tailor the gain and output range to a specific application.
The entire electronics package fits into a compact Eurocard rack module (with the high-voltage section taking only a fraction of the PCB), and draws less than \SI{150}{\milli\ampere} at \SI{15}{\volt}.
The high-voltage output current will be limited by the switching regulator and by the LM7171 op-amp used for low-noise stabilization (U2 in Fig.~\ref{Fig:PiezoCircuit}, which can supply at most $\approx\SI{100}{\milli\ampere}$\cite{LM7171Datasheet}), but is sufficient for nearly all piezoelectric applications.

\begin{figure}[t!]
\includegraphics[width=\columnwidth]{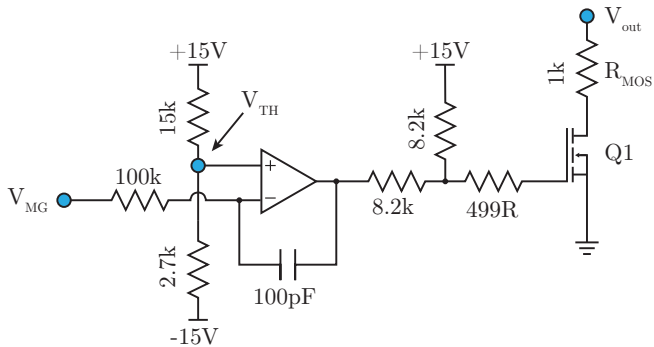}
\caption{MOSFET ``quench'' circuit.
When the mid-ground node $\text{V}_\text{MG}$ (also shown in Fig.~\ref{Fig:PiezoCircuit}) goes below the threshold value set at $\text{V}_\text{TH}$, the op-amp puts the HV MOSFET Q1 into conduction.
When engaged, the quench time constant is given by $\tau \approx R_\text{MOS} \text{C}_\text{out}$, which for our circuit is set to \SI{1}{\milli\second}. The capacitor $\text{C}_\text{out}=\SI{1}{\micro\farad}$ is shown in Fig.~\ref{Fig:PiezoCircuit}, and details of this circuit are discussed in Sec.~\ref{Sec:SlowModulationMOS}.
\label{Fig:MOS}}
\end{figure}

Fig.~\ref{Fig:PiezoCircuit} shows an overview of the circuit schematic. 
A flyback regulator (Sec.~\ref{Sec:DRV2700}) controls the potential between the high-voltage ($\text{V}_\text{HV}$) and floating-ground ($\text{V}_\text{FG}$) circuit nodes, while the low-noise stabilization circuitry (Sec.~\ref{Sec:LowNoiseStabilization}) controls the output node $\text{V}_\text{out}$ relative to the true circuit ground.
A digital-to-analog converter (DAC) generates a voltage setpoint, $\text{V}_\text{ctl}$, which is sent to the high-voltage flyback regulator and to the low-noise stabilization circuit.
The DAC is controlled by an integrated microcontroller, and can be programmed to output slow ($\approx\SI{10}{\hertz}$) rail-to-rail voltage ramps in addition to setting the DC operating point (Sec.~\ref{Sec:SlowModulationMOS}).
To improve the large-amplitude slew rate, a MOSFET ``quench'' circuit (Fig.~\ref{Fig:MOS}) is included to reduce the $RC$ time constant of the high-voltage node $\text{V}_\text{out}$ when needed.
A low-pass filter with a switchable corner frequency can be engaged during DC operation to reduce $1/f$ noise from the DAC (discussed in Sec.~\ref{Sec:NoiseAnalysis}).
Fast output modulation between $\pm\SI{10}{\volt}$ can be achieved through the input node $\text{V}_\text{mod}$.
This node is DC-coupled to the high-voltage output, and is useful for closed-loop feedback control.

\subsection{Flyback regulator}
\label{Sec:DRV2700}

The high-voltage DC-DC converter used here is based on the Texas Instruments DRV2700 piezo driver.\footnote{The identification of commercial products in this paper is for information only and does not imply recommendation or endorsement by the National Institute of Standards and Technology.}
This integrated circuit can be operated as a boost converter to drive an on-chip differential amplifier up to \SI{100}{\volt}, or as a flyback converter up to \SI{1}{\kilo\volt} or more.
In flyback configuration, the internal-boost switch of the DRV2700 (pin SW in Fig.~\ref{Fig:PiezoCircuit}) drives a step-up transformer.
When the switch closes, current begins to flow through the primary coil of the transformer and induces a corresponding voltage across the secondary coil.
In this state, the output diode D1 is reverse-biased, and the capacitor ($\text{C}_{\text{HV}}$ in Fig.~\ref{Fig:PiezoCircuit}) holds its charge.
When the switch opens, the voltage across the secondary coil is inverted, putting the diode into conduction and charging the capacitor.
By changing the switching duty cycle, the DRV2700 is able to regulate the voltage across the galvanically isolated output (nodes $\text{V}_\text{FG}$ and $\text{V}_\text{HV}$ in Fig.~\ref{Fig:PiezoCircuit}).

The DRV2700 implements output voltage control by comparing the feedback input pin at node $\text{V}_\text{FB}$ with an internal (\SI{1.3}{\volt}) reference.
The resistors $R_9$ and $R_{10}$ are chosen such that pin FB is at \SI{1.3}{\volt} when the output of U1 is at ground: $R_{10}/(R_9+R_{10}) = \SI{1.3}{\volt}/\SI{5}{\volt} \approx \num{0.26}$.
The op-amp U1 subtracts $\text{V}_\text{FG}$ and $G\cdot \text{V}_\text{ctl}$ from the voltage at node $\text{V}_\text{HV}$, ensuring the DRV2700 regulates the output voltage such that
\begin{align}
\label{Eq:U1Output}
\text{V}_\text{HV} - \text{V}_{\text{FG}} &= G\cdot \text{V}_{\text{ctl}}\,,
\end{align}
where the gain $G$ is set by the resistor ratio $R_3/R_4 \equiv R_5/R_6$, and $\text{V}_\text{ctl}$ is the control voltage set by the DAC.
The capacitors $C_3$ and $C_4$ are chosen such that $C_3 = \SI{22}{\pico\farad}$ and
\begin{align}
\frac{C_4}{C_3} &= \frac{R_3}{R_4~||~R_7}\,,
\end{align}
as suggested by the DRV2700 datasheet\cite{DRV2700Datasheet}, where $R_6 = R_7 = R_8$, and $R_i~||~R_j \equiv R_i R_j/(R_i + R_j)$.
In our implementation, we choose a gain $G\approx 50$ ($R_3 = R_5 = \SI{499}{\kilo\ohm}$; $R_4 = R_{6,7,8} = \SI{10}{\kilo\ohm}$), allowing a \SI{5}{\volt} control signal $\text{V}_\text{ctl}$ to span \SI{250}{\volt} at the output. 
A different DAC and/or a different gain factor could be chosen to adjust the maximum output voltage.\footnote{Note that the passive components must be rated for the chosen output voltage.}
The transformer (ATB3225, 1:10 step-up winding), diode, and RC feedback network are all based on values suggested in the DRV2700 datasheet.\cite{DRV2700Datasheet,DRV2700EVMUserGuide}

The output of the flyback regulator is passed through a four-pole, low-pass RC filter.
The corner frequency $f_c \approx \SI{30}{\hertz}$ is chosen to be high enough that a slow ($\approx \SI{10}{\hertz}$) rail-to-rail triangle ramp can be applied by the DAC at $\text{V}_\text{ctl}$ (for, e.g., sweeping over a resonance in spectroscopy), but low enough that the $\approx \SI{100}{\kilo\hertz}$ switching noise is substantially attenuated.
Additional capacitors on both the $\text{V}_\text{HV}$ and $\text{V}_\text{FG}$ resistor networks shunt high frequency noise to ground.
This filter topology, modeled on a lossy transmission line, is sufficient for our application, but other corner frequencies or topologies could also be used.

%% place figure here for better text flow

\subsection{Low-noise stabilization and fast modulation}
\label{Sec:LowNoiseStabilization}

The low-noise stabilization circuit is crucial to the performance of the design, as it is responsible for removing noise at the output of the flyback converter.
To accomplish this, a high slew-rate op-amp (Texas Instruments LM7171, \SI[per-mode=symbol]{4100}{\volt\per\micro\second}; see U2 in Fig.~\ref{Fig:PiezoCircuit}) drives the galvanically isolated floating ground of the flyback converter.
This op-amp senses the voltage $\text{V}_\text{out}$ referenced to true circuit ground, and adjusts its output such that
\begin{align}
\text{V}_\text{out} &= \left(1 + \frac{R_1}{R_2~||~R_\text{mod}}\right) \text{V}_\text{ctl} -
\left(\frac{R_1}{R_\text{mod}}\right) \text{V}_\text{mod}\,.
\label{Eq:FullTransferFunc}
\end{align}
Here, $\text{V}_\text{mod}$ is the applied modulation, which can vary between $\pm\SI{10}{\volt}$ and is inverted before being summed onto the output.
We choose $R_1 = R_\text{mod} = \SI{1}{\mega\ohm}$ and $R_2 = \SI{20.5}{\kilo\ohm}$ such that the DC gain $\Delta\text{V}_\text{out}/\Delta\text{V}_\text{ctl}$ is $\approx 50$ and the modulation gain $\Delta\text{V}_\text{out}/\Delta\text{V}_\text{mod}$ is unity.
Depending on the application, other gain configurations could work equally well provided the non-inverting gain of U2 closely matches the gain of the flyback regulator (since they both derive from $\text{V}_\text{ctl}$).

The op-amp U2 controls $\text{V}_\text{out}$ via two different feedback pathways.
At low frequencies, it modifies the floating ground reference $\text{V}_\text{FG}$ of the flyback converter, and the DRV2700 in turn modifies $\text{V}_\text{HV}$ to maintain a constant voltage between $\text{V}_\text{FG}$ and $V_\text{HV}$.
At higher frequencies, U2 is decoupled from $\text{V}_\text{FG}$ by the passive filtering network.
In this regime, $\text{C}_\text{out}$ provides a low-impedance path between U2 and the output, such that high-frequency switching noise can be directly compensated for by the op-amp.
We chose a value $\text{C}_\text{out}=\SI{1}{\micro\farad}$, which is a compromise between component size and the desire for a large capacitance.
In addition, a small resistance $R_{11} = \SI{50}{\ohm}$ is inserted between U2 and $\text{C}_\text{out}$ to ensure stable operation.
Smaller $R_{11}$ and/or larger $\text{C}_\text{out}$ might provide better performance, but this has not been tested.

The choice of components for resistors $R_1$ and $R_2$ is crucial for the low-noise performance of the system.
Because this resistive divider is responsible for accurately sensing the voltage $\text{V}_\text{out}$, noise introduced by these resistors cannot be corrected by the op-amp.
In general, resistors are fundamentally limited by Johnson noise, in which thermal fluctuations contribute to a white noise power spectrum given by $4 k_B T R$, where $T$ is the temperature and $k_B$ is Boltzmann's constant.\cite{Horowitz2015a:JN}
However, resistors also exhibit $1/f$ current noise caused by equilibrium fluctuations of the resistance.\cite{Clarke1974a,Voss1976a}
This excess noise depends on the applied voltage, and therefore is an important consideration in a high-voltage circuit.
It is also highly dependent on the resistor composition and varies from manufacturer to manufacturer.
Seifert, et. al.~\cite{Seifert2009a} characterized $1/f$ noise in a variety of resistors, and found that the Vishay TNPW \SI{0.1}{\percent}-series resistors showed a noise spectrum almost consistent with Johnson noise down to \SI{1}{\hertz}.
Our current design uses this series in a 1206 package, but we noticed low-frequency noise correlated with varying strain on the PCB, potentially due to the relatively large footprint of this package.
Future boards might instead use three TNPW \SI{0.1}{\percent} 0603 resistors in series for both R1 and R2 to minimize strain-induced output noise.

The value of capacitor $\text{C}_1$ is a tradeoff between two competing design considerations.
On the one hand, a larger $\text{C}_1$ extends the frequency range where switching noise from the DRV2700 is suppressed.
However, large values of $\text{C}_1$ limit the bandwith of $\text{V}_\text{ctl}$.
We empirically settled on $\text{C}_1 = \SI{1}{\nano\farad}$, which is large enough to saturate the feedback gain in the \SI{40}{\kilo\hertz}--\SI{100}{\kilo\hertz} range where switching noise dominates, but not so large that it limits the bandwidth of $\text{V}_\text{ctl}$ below the corner set by the switchable low-pass filter described in Sec.~\ref{Sec:SlowModulationMOS}.
Once $\text{C}_1$ was chosen, capacitor $\text{C}_\text{mod}$ was calculated to match the impedances $R_1~||~\text{C}_1 = R_\text{mod}~||~\text{C}_\text{mod}$.
For our circuit, this means $\text{C}_\text{mod} = \text{C}_1$.

\subsection{Digital control and slow modulation}
\label{Sec:SlowModulationMOS}

The high-voltage setpoint (absent voltages summed in at $\text{V}_\text{mod}$) is controlled by a low-noise DAC.
This has two advantages: digital control enhances setpoint repeatability, and simplifies the integration with computerized control electronics or sophisticated servo loops.
While the modulation input $\text{V}_\text{mod}$ has a limited range of $\pm\SI{10}{\volt}$, larger voltage swings can be achieved by reprogramming the DAC.

As discussed below in Sec.~\ref{Sec:NoiseAnalysis}, without modification the DAC would dominate the noise performance of $\text{V}_\text{out}$.
Therefore we add a single-pole, low-pass filter between $\text{V}_\text{ctl}$ and the non-inverting node of U2 to bring the DAC noise contribution below other noise sources in the circuit.
This filter has a switchable corner frequency (between \SI{165}{\hertz} and \SI{0.8}{\hertz}) to optimize noise performance at DC while still permitting AC modulation when needed.
It consists of a \SI{20.5}{\kilo\ohm} resistor and \SI{47}{\nano\farad} capacitor, with a secondary \SI{10}{\micro\farad} capacitor that can be switched in to operate with the lower corner frequency.

One downside of the flyback regulator presented above is that while the switched transformer can quickly charge the output capacitors, the discharge time $\tau$ is limited to {$\approx~\SI{1}{\second}$} by the RC time constant of the circuit.
To get around this limitation, we have added an auxiliary MOSFET ``quench'' circuit\footnote{The quench circuit presented here is based on the pulldown FET discussed in the DRV2700 datasheet, with some additional modifications to suit our purposes.} to quickly shunt $\text{V}_\text{out}$ to ground (see Fig.~\ref{Fig:MOS}).
This circuit works by monitoring the voltage at $\text{V}_\text{MG}$, the mid-ground node controlled by op-amp U2.
If $\text{V}_\text{MG}$ drops below a threshold set by $\text{V}_\text{TH}$, the comparator op-amp in Fig.~\ref{Fig:MOS} changes the gate voltage of the MOSFET to put it into conduction.
The time constant for this configuration is given by $\tau \approx R_\text{MOS}\text{C}_\text{out}$.
For our circuit, this changes $\tau$ to $\approx\SI{1}{\milli\second}$, allowing $\text{C}_\text{out}$ to be quickly discharged.
The threshold is $\text{V}_\text{TH} = \SI{-10.4}{\volt}$, but other values could be chosen depending on the design requirements.

The high-voltage design presented here has the flexibility to exist as a standalone circuit or be integrated with other electronics, and we have included several auxiliary features to make this more convenient.
For example, the analog modulation input is differentially buffered to break ground loops (not shown in Fig.~\ref{Fig:PiezoCircuit}), and the digital portion can be interfaced with other devices in the lab to expand conceivable control scenarios.
Secondary features include a divided-down output that can be used as a monitor or fed forward to a low-noise laser diode current controller like the one in Erickson, et. al.~\cite{Erickson2008a}
Of course many variations are possible, and we refer the reader to our GitHub page for more details on our specific implementation.\cite{PiezoDesignFiles}

\section{Noise model \& analysis}
\label{Sec:NoiseAnalysis}

\begin{figure}[t]
\includegraphics[width=\columnwidth]{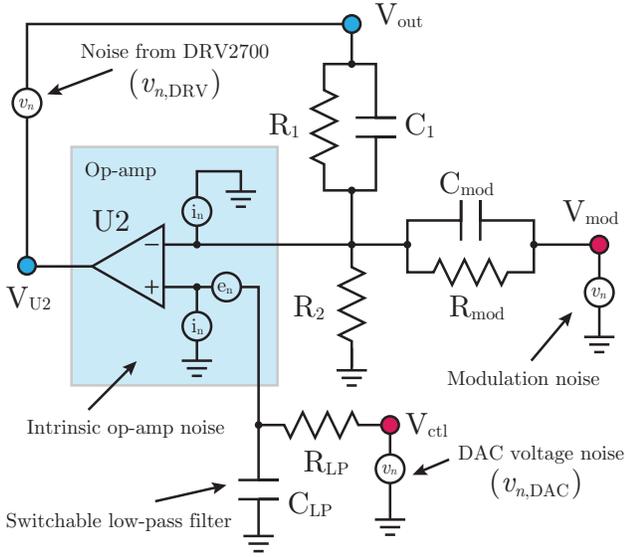}
\caption{Noise model, including contributions from the op-amp, DAC, DRV2700, and modulation input. The element labeled $\text{C}_\text{LP}$ consists of a \SI{47}{\nano\farad} capacitor in parallel with a \SI{10}{\micro\farad} capacitor connected to ground through a switch, such that the corner frequency of the filter can be changed depending on the mode of operation  (see text). Though not drawn in the figure, we also consider the Johnson noise contributions from all resistors.
\label{Fig:NoiseModel}}
\end{figure}

\begin{figure}[t]
\subfloat{
    \includegraphics[width=\columnwidth]{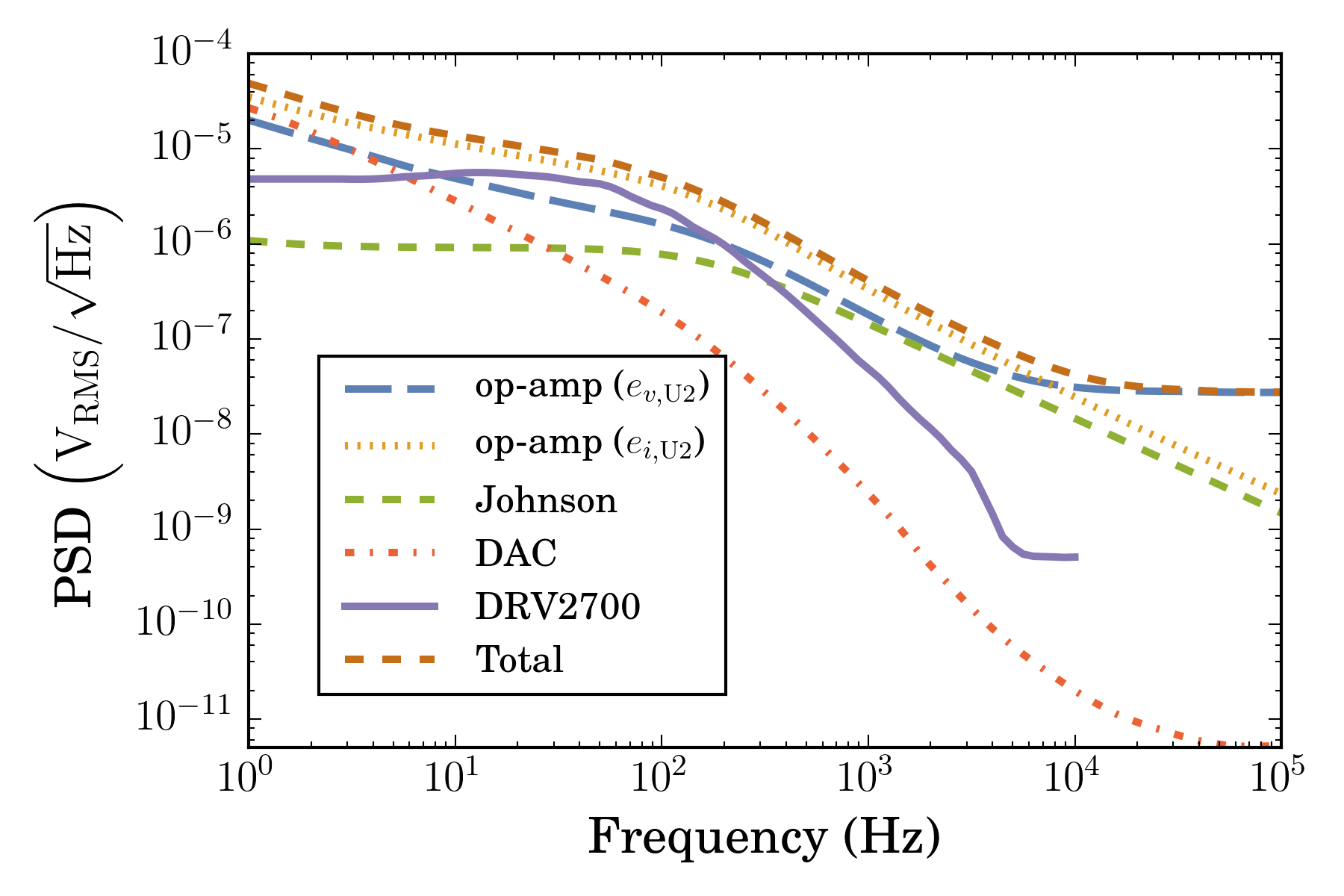}
}

\subfloat{
    \centering
    \begin{tabular*}{\columnwidth}{@{\extracolsep{\fill}} llcc}
    \hline\hline
    \rule{0pt}{2.5ex}\multirow{2}{*}{Noise source} & & RMS Voltage & RMS Voltage \\
    & & (\SI{1}{\hertz}~--~\SI{10}{\hertz})\rule[-1ex]{0pt}{1ex} & (\SI{10}{\hertz}~--~\SI{100}{\kilo\hertz})\rule[-1ex]{0pt}{1ex} \\
    \hline
    \rule{0pt}{2ex}op-amp voltage & ($e_{v,\text{U2}}$)  & \SI{26}{\micro\volt} & \SI{31}{\micro\volt}      \\
    op-amp voltage &  ($e_{i,\text{U2}}$)  &            \SI{51}{\micro\volt}     & \SI{73}{\micro\volt}      \\
    DAC             &  ($e_{n,\text{DAC}}$)  &          \SI{28}{\micro\volt}    & \SI{8}{\micro\volt}       \\
    Johnson-Nyquist &   ($e_{n,\text{JN}}$)  &         \SI{3}{\micro\volt}    & \SI{14}{\micro\volt}        \\ 
    DRV2700 & ($e_{n,\text{DRV}}$) &        \SI{15}{\micro\volt}   & \SI{43}{\micro\volt} \\
    \hline
    \rule{0pt}{2ex}\textbf{total (calculated)} & & \SI{66}{\micro\volt} & \SI{92}{\micro\volt}  \\
    \hline\hline
    \end{tabular*}
}
\caption{Modeled noise contributions (color online). Plotted are the noise contributions from each source in our model, along with the total calculated noise. Power spectral density (PSD) is referred to the high-voltage output, and the table shows the integrated RMS noise due to each noise source in different frequency bands. The total RMS noise (summed in quadrature) over the entire \SI{1}{\hertz}~--~\SI{100}{\kilo\hertz} range is calculated to be \SI{113}{\micro\volt}, with the residual DRV2700 switching noise measured at \SI{100}{\volt} as described in the text. Frequency-dependent noise spectra for the DAC and op-amp were extracted from the datasheets.\label{Fig:NoisePlot}}
\end{figure}

To better understand the circuit performance and the measured output noise reported in Sec.~\ref{Sec:Results}, we introduce the noise model shown in Fig.~\ref{Fig:NoiseModel}. 
A summary of each noise contribution (op-amp, DAC, Johnson-Nyquist, and residual ripple from the DRV2700, all calculated at the node $\text{V}_\text{out}$) is shown in Fig.~\ref{Fig:NoisePlot}, along with the cumulative root-mean-square (RMS) noise estimates in different frequency bands.
We will neglect noise appearing at node $\text{V}_\text{mod}$ due to the external modulation because its exact character depends on the external drive.

To facilitate the noise analysis, we calculate the voltage and current (transimpedance) gains from the input nodes of U2 to the output, $\text{V}_\text{out}$.
Starting with the non-inverting node, we find the voltage gain to be
\begin{align}
G_{+}^{(v)} = 1 + Z_1\left(\frac{1}{R_2} + \frac{1}{Z_\text{mod}}\right)\,,
\end{align}
where $Z_1$ and $Z_\text{mod}$ are the equivalent impedances of $R_1~||~\text{C}_1$ and $R_\text{mod}~||~\text{C}_\text{mod}$, respectively.
The transimpedance gain, $G^{(i)}_{+}$, follows by multiplying $G_+^{(v)}$ by the impedance, $Z_+$, seen from that node. % $Z_+ = R_\text{LP}~||~\text{C}_\text{LP}$ seen from that node.
Thus,
\begin{align}
G_{+}^{(i)} = G_{+}^{(v)}\left[\frac{R_\text{LP}}{1+2\pi \iu f R_\text{LP} C_\text{LP}}\right]\,,
\end{align}
where the bracketed term is $Z_+$, $f$ is the Fourier frequency, and $\iu$ is the imaginary unit.

We now calculate gains from the inverting node of U2.
Because any currents appearing at this node will be cancelled by the feedback of U2, the transimpedance gain $G_{-}^{(i)}$ is simply the impedance $Z_1$, given by
\begin{align}
G_{-}^{(i)} = Z_1 \equiv \frac{R_1}{1+2\pi \iu f R_1 C_1}\,.
\end{align}
With these expressions in hand, we can calculate the output noise contribution from each source in our model.

As shown in Fig.~\ref{Fig:NoiseModel}, the op-amp noise is parametrized by two noise contributions: $e_n$, the input voltage noise spectral density, and $i_n$, the input current noise spectral density.
For the LM7171 at $\SI{10}{\kilo\hertz}$, $e_n = \SI[per-mode=symbol]{14}{\nano\volt\per\sqrthz}$ and $i_n = \SI[per-mode=symbol]{1.5}{\pico\ampere\per\sqrthz}$ with a $1/f$ noise character below this frequency.\cite{LM7171Datasheet}
The voltage noise appears at the non-inverting input, while the current noise is present at both inputs.
By multiplying each source by the appropriate gain, we obtain the equivalent output-noise power spectral densities (PSD),
\begin{align}
\begin{split}
e^2_{v,\text{U2}} &= \big|G_{+}^{(v)}\big|^2 e^2_n \\
e^2_{i,\text{U2}} &= \big|G_{-}^{(i)}\big|^2 i^2_n + \big|G_{+}^{(i)}\big|^2 i^2_n\,,
\end{split}
\end{align}
where  $e^2_{v,\text{U2}}$, $e^2_{i,\text{U2}}$ are the output-referred voltage and current noise PSD, respectively (plotted in Fig.~\ref{Fig:NoisePlot}).

Next, we calculate the DAC noise contribution.
The voltage gain from the node $\text{V}_{\text{ctl}}$ is given by
\begin{align}
\label{Eq:Gdc}
G^{(v)}_{\text{ctl}} &= \left[\frac{1}{1+2\pi \iu f R_\text{LP} C_\text{LP}}\right]G_{+}^{(v)}\,,
\end{align}
where the bracketed term represents the contribution to the transfer function from the switchable low-pass filter.
Without the addition of this low-pass filter, the voltage noise of the DAC would dominate both the low- and high-frequency noise performance of the circuit.
A simple solution would be to place a fixed-corner filter at this node, but this would severely restrict the AC performance of $\text{V}_\text{ctl}$.
Thus, we use a switchable low-pass filter (as described in Sec.~\ref{Sec:SlowModulationMOS}) to achieve low-noise performance during DC operation, while still permitting the DAC to modulate $\text{V}_\text{ctl}$ more quickly when needed.
The non-zero resistance of the switch contributes a zero to the transfer function at $\approx\SI{23}{\kilo\hertz}$, but has negligible effect on the performance.
The DAC voltage noise contribution can now be calculated as 
\begin{align}
e^2_{n,\text{DAC}} = \big|G^{(v)}_{\text{ctl}}\big|^2 v^2_{n,\text{DAC}}\,,
\end{align}
where $v_{n,\text{DAC}}$ is the frequency-dependent output voltage noise of the DAC as reported in the datasheet.\cite{AD56XXRDatasheet} 
In subsequent calculations, we take the DC-mode operation ($f_c = \SI{0.8}{\hertz}$) for the switchable low-pass filter.
 
The Johnson noise contribution can be calculated by modeling each resistor with a parallel current noise given by $i_\text{R}^2 = 4k_B T/R$.
Resistors $R_1$, $R_2$, and $R_\text{mod}$ all contribute current noise at the inverting node of U2, which as discussed previously has a transimpedance gain to the output given by $G_{-}^{(i)}$.
The resistor $R_\text{LP}$ contributes current noise at the non-inverting node, which sees a transimpedance gain $G_{+}^{(i)}$.
Thus, the total Johnson noise is
\begin{align}
\begin{split}
e^2_{n,\text{JN}} &= 4 k_B T \times \\
&\left[ \big|G_{-}^{(i)}\big|^2 \left(\frac{1}{R_1} + \frac{1}{R_2} + \frac{1}{R_\text{mod}} \right) + 
\frac{\big|G_{+}^{(i)}\big|^2}{R_\text{LP}} \right]\,.
\end{split}
\end{align}

The low-noise stabilization circuit is limited in its ability to reject residual switching noise from the DRV2700 regulator (after the passive filtering network) by the total loop gain analyzed from node $\text{V}_\text{out}$.
The LM7171 has a reported open-loop gain of \SI{85}{\decibel} ($\approx 1.8\times 10^4$), with a dominant pole at $\approx\SI{10}{\kilo\hertz}$.\cite{LM7171Datasheet}
We can model the open loop gain, $G_\text{OL}$, as
\begin{align}
G_\text{OL} &= \frac{1.8\times10^4}{1 + \iu (f/\SI{10}{\kilo\hertz})}\,.
\end{align}
The feedback network contributes a gain
\begin{align}
G_\text{FB} = \frac{R_2~||~Z_\text{mod}}{Z_1 + R_2~||~Z_\text{mod}}\,,
\end{align}
arising from the voltage partition between $\text{V}_\text{out}$ and the inverting node of U2.
From these, we write down the closed-loop gain seen from $\text{V}_\text{out}$,
\begin{align}
G^{(\text{DRV})} = \frac{1}{1+G_\text{OL} G_\text{FB}}\,.
\end{align}
The contribution to the output from residual switching noise, $v_{n,\text{DRV}}$, is then
\begin{align}
e^2_{n,\text{DRV}} = v^2_{n,\text{DRV}}\big|G^{(\text{DRV})}\big|^2\,.
\end{align}
For $R_1=\SI{1}{\mega\ohm}$ and $\text{C}_1 = \SI{1}{\nano\farad}$, $v_{n,\text{DRV}}$ is attenuated by as much as \SI{76}{\decibel} at \SI{6.3}{\kilo\hertz}.

We estimate the noise spectral density $v_{n,\text{DRV}}$ by monitoring the node $\text{V}_\text{U2}$ in Fig.~\ref{Fig:NoiseModel}, since the output of this op-amp represents the control signal required to cancel voltage fluctuations at $\text{V}_\text{out}$.
The trace for $e_{n,\text{DRV}}$ plotted in Fig.~\ref{Fig:NoisePlot} is derived from the results of this measurement.
Because the measured $v_{n,\text{DRV}}$ drops below the noise floor of our spectrum analyzer at $\approx \SI{10}{\kilo\hertz}$, we only plot the trace out to this frequency.

\begin{figure}[b]
\includegraphics[width=\columnwidth]{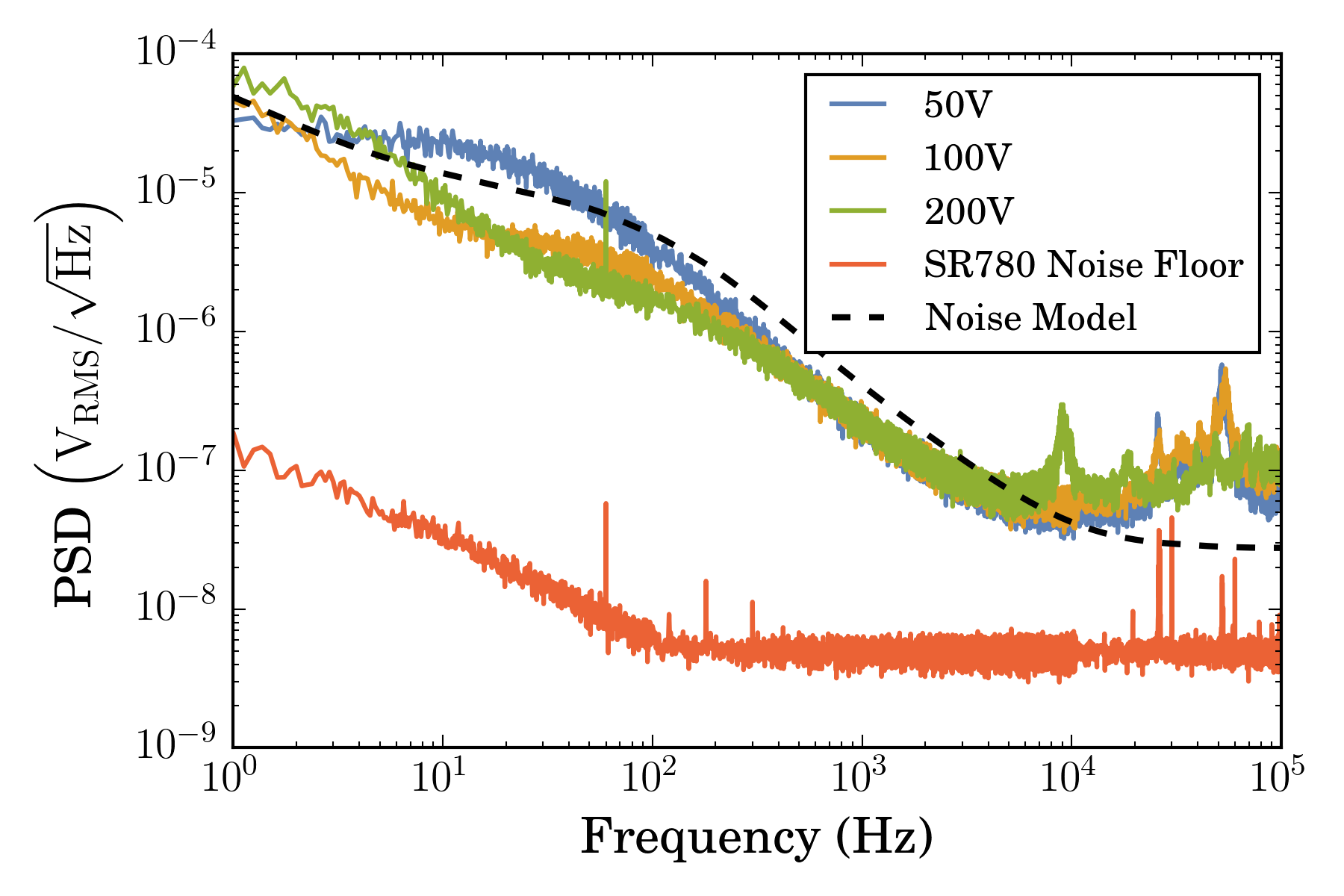}
\caption{Voltage noise power spectral density at various output voltages, as measured on an SR780 spectrum analyzer (color online). The integrated RMS noise ($\SI{1}{\hertz} - \SI{100}{\kilo\hertz}$) is $\{138, 80, 101\}~\si{\micro\volt}$ measured at  $\{50, 100, 200\}~\si{\volt}$, with no output load. The dashed black line overlays the noise estimate from our model, calculated at \SI{100}{\volt} (see Fig.~\ref{Fig:NoisePlot}).\label{Fig:PSD}}
\end{figure}

Given the noise model discussed above, our circuit is dominated by the op-amp's intrinsic current noise at lower frequencies, and voltage noise at higher frequencies.
The op-amp current noise contribution could be suppressed by using lower resistances $R_1$ and $R_\text{mod}$, however one must be careful about power and current limitations when dealing with such high voltages.
Each noise source is tabulated and plotted in Fig.~\ref{Fig:NoisePlot}, and the total voltage noise ($\SI{1}{\hertz} - \SI{100}{\kilo\hertz}$, $\text{V}_\text{out} = \SI{100}{\volt}$) is calculated to be $\SI{113}{\micro\volt}_\text{RMS}$ .

\section{Results}
\label{Sec:Results}

\begin{figure}[b]
\includegraphics[width=\columnwidth]{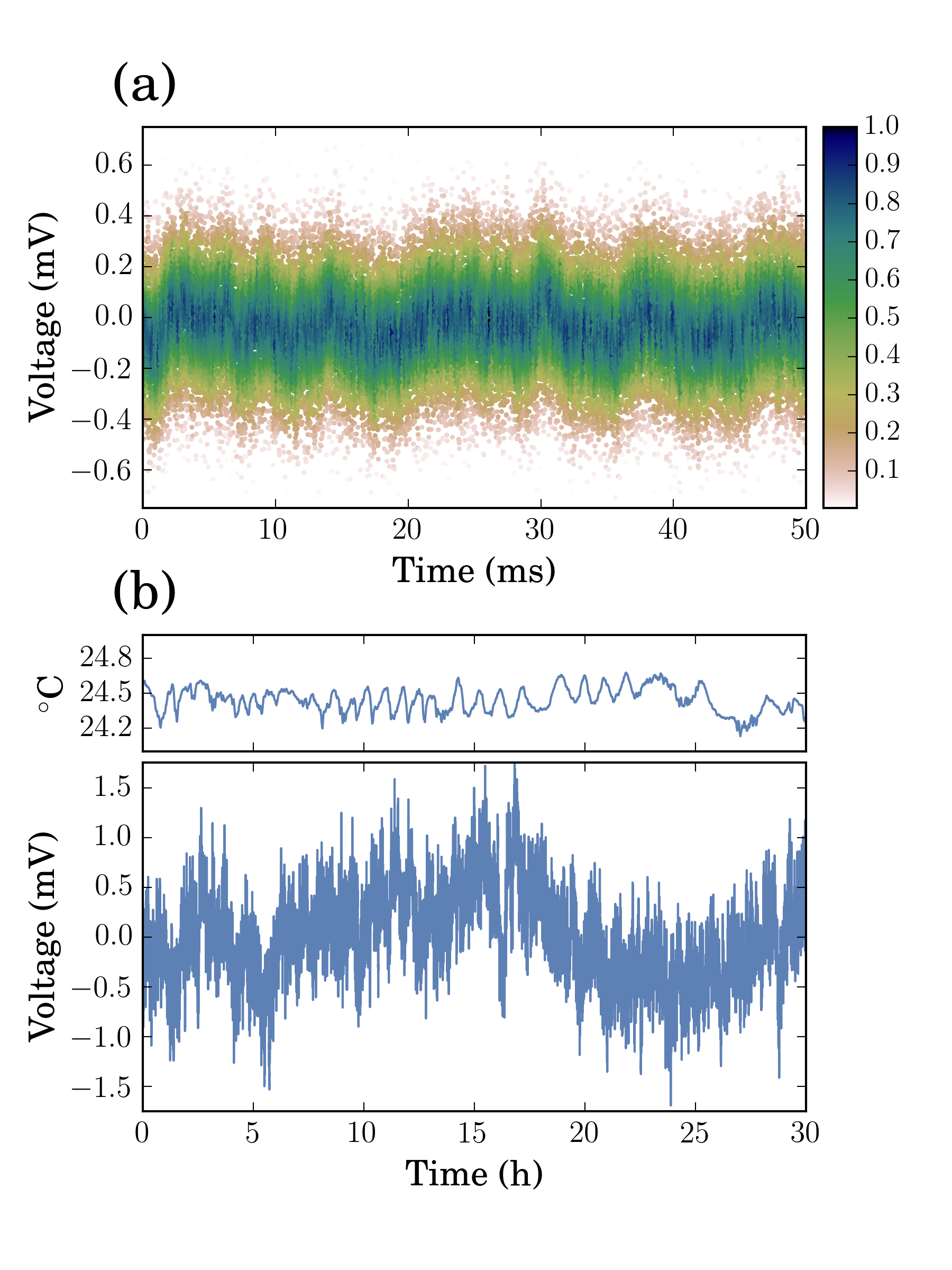}
\caption{Time-domain traces of the high-voltage output (\SI{100}{\volt}, color online). (a) Short-time scatterplot, measured on a PicoScope~5442B (AC coupled).
The points have been down-sampled for clarity, and colored based on a normally-distributed probability of occurence in each 100-\si{\micro\second} timeslice (scaled to the most probable voltage).
The color thus provides a visual estimation of the RMS width.
(b) Long-term trace, measured on a Keithly~2010 digital multimeter. A~\SI{100}{\volt} DC offset is subtracted from the plotted values. The top panel shows the lab temperature during the same time period.\label{Fig:TimeDomain}}
\end{figure}
 
The measured performance of the high-voltage piezo driver is shown in Fig.~\ref{Fig:PSD}, where we plot the noise power spectral density measured at several different output voltages.
These traces were taken on a Stanford Research Systems SR780 spectrum analyzer, with the high-voltage output coupled through a \SI{0.5}{\hertz} high-pass filter and without any capacitive load.
This represents a worst-case scenario, as larger capacitances at the output will reduce the noise.
The integrated noise ($\SI{1}{\hertz} - \SI{100}{\kilo\hertz}$) was measured to be $\{\SI{138}{\micro\volt}_\text{RMS},~\SI{80}{\micro\volt}_\text{RMS},~\SI{101}{\micro\volt}_\text{RMS}\}$ for \{\SI{50}{\volt}, \SI{100}{\volt}, \SI{200}{\volt}\} outputs. 
This matches roughly with the expected total RMS noise calculated in Sec.~\ref{Sec:NoiseAnalysis}.
The difference in noise performance at different output voltages can be traced back to the residual ripple of the DRV2700, which is larger for lower output voltages.
Indeed, one can see the characteristic shape change in Fig.~\ref{Fig:PSD} between \SI{100}{\volt} and \SI{50}{\volt} as the residual ripple $e_{n,\text{DRV}}$ begins to dominate at low frequencies.

\begin{figure}[t]
\includegraphics[width=\columnwidth]{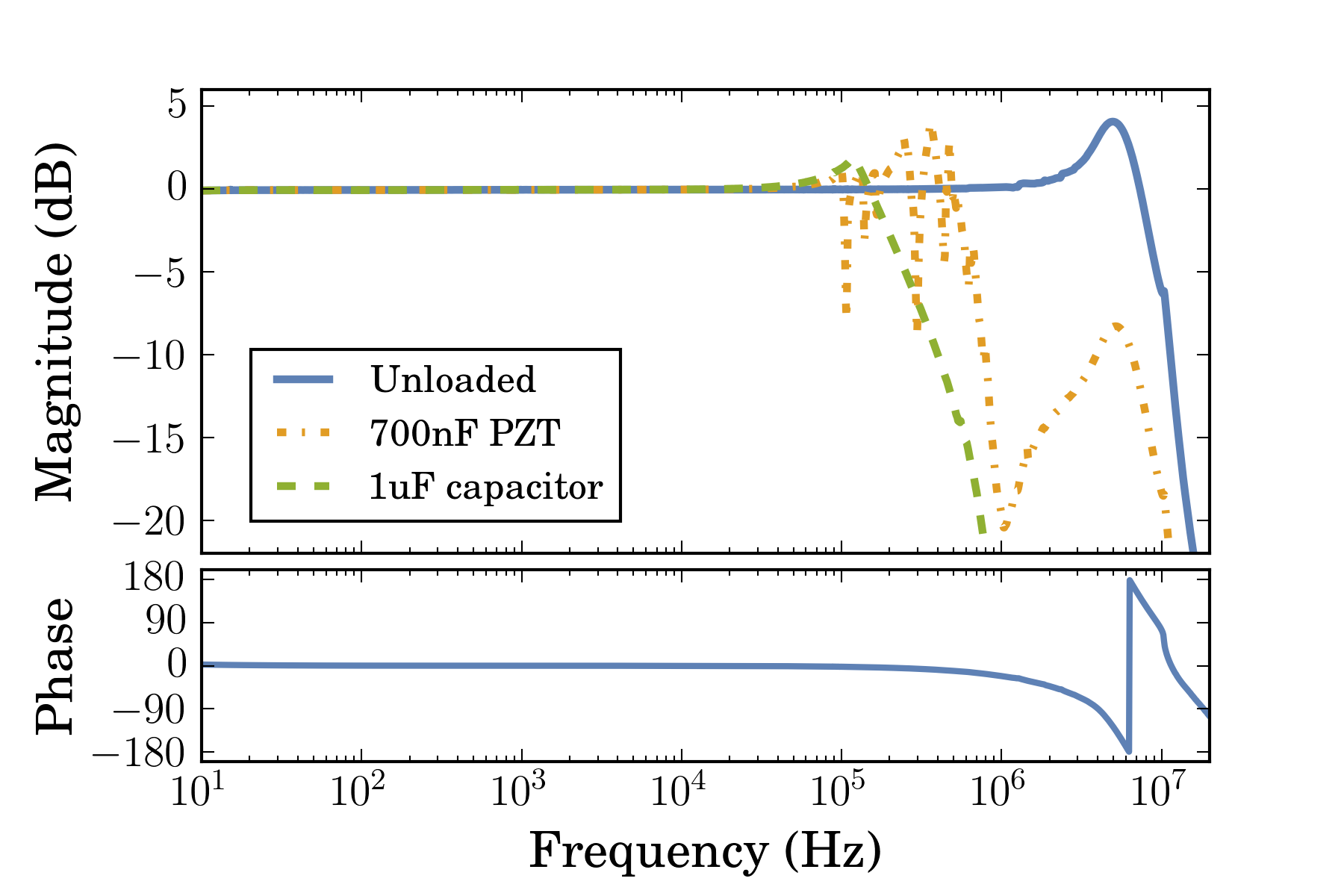}
\caption{Modulation input transfer function (color online). The solid blue line indicates the unloaded frequency response, with the phase plotted in the lower panel. The dash-dotted orange trace shows the response with a \SI{700}{\nano\farad} piezoelectric actuator~{(Thorlabs PN~AE0505D08F)}. Several mechanical resonances above $\approx\SI{50}{\kilo\hertz}$ are clearly visible. The dashed green trace shows the response under a \SI{1}{\micro\farad} capacitive load. Loaded response bandwidth is $\approx~\SI{100}{\kilo\hertz}$, and the unloaded gain is flat within $\SI{0.1}{\decibel}$ out to $\SI{1}{\mega\hertz}$ where the phase is \SI{-20}{\degree}.\label{Fig:TransferFunc}}
\end{figure}

Fig.~\ref{Fig:TimeDomain} shows the performance at both short- and long-time scales.
At long times, voltage fluctuations on the order of a few \si{\milli\volt} can be observed.
This is due generically to $1/f$ noise, but also correlates with the external temperature.
A cross-correlation between the measured temperature and output voltage yields an effective temperature coefficient of \SI[per-mode=symbol]{-24}{ppm\per\celsius} at \SI{100}{\volt}.\footnote{This value was extracted from a separate long-term voltage measurement, during which the temperature of the room fluctuated by $\approx \SI{3}{\celsius}$.}
The short-term trace was taken on a PicoScope~5442B (AC-coupled, \SI{100}{\volt} output), and downsampled for clarity.
Points are binned into 100-\si{\micro\second} slices, and colored based on their normally-distributed probability of occurrence.
The color scale is normalized to the most probable voltage in each bin.

Fig.~\ref{Fig:TransferFunc} shows the measured frequency response under different load conditions. The unloaded bandwidth is as high as a few megahertz, while a \SI{1}{\micro\farad} capacitive load can still be driven at $\approx\SI{100}{\kilo\hertz}$. 
Several mechanical resonances can be seen with a \SI{700}{\nano\farad} piezoelectric load, as expected.
In a laboratory setting, these resonances can be mitigated by using a digital feedback controller with notch filters tuned to match the exact resonance frequencies observed in the system\cite{Ryou2016a}, thereby extending the usable bandwidth out to $\approx\SI{100}{\kilo\hertz}$.

\section{Conclusion}
\label{Sec:Conclusion}

We have designed, built, and characterized a high-voltage piezoelectric driver optimized for use in a modern atomic physics laboratory.
It is based on a flyback configuration switching regulator, but is able to achieve very low noise performance by active stabilization from a high slew-rate op-amp.
This hybrid architecture makes it small and easy to deploy in a variety of situations, without requiring an external high-voltage power supply.
The design principles discussed here can be adapted to fit the exact application, and all design files are freely available on GitHub for others to use and modify.

\begin{acknowledgments}
The authors would like to thank Z. Smith and D. Genkina for useful discussions.
This work was partially supported by the Office of Naval Research, and the National Science Foundation through the Physics Frontier Center at the Joint Quantum Institute.
\end{acknowledgments}

\bibliography{references}

\end{document}